\documentclass[prb,citeautoscript,twocolumn,amsmath,amssymb]{revtex4}
\usepackage{bm}
\usepackage{graphicx}

\begin{document}


\newcommand{\dtwomin}{D^2_\text{min}}
\newcommand{\eg}{\emph{e.g.}\ }
\newcommand{\ie}{\emph{i.e.}\ }

\title{Role of disorder in finite-amplitude shear of a 2D jammed material}

\author{Nathan~C.~Keim}
\email{nkeim@calpoly.edu}
\affiliation{Department of Mechanical Engineering and Applied Mechanics, University of Pennsylvania, Philadelphia, PA 19104}
\affiliation{Department of Physics, California Polytechnic State University, San Luis Obispo, CA 93407}
\author{Paulo~E.~Arratia}
\email{parratia@seas.upenn.edu}
\affiliation{Department of Mechanical Engineering and Applied Mechanics, University of Pennsylvania, Philadelphia, PA 19104}

\date{\today}

\begin{abstract}

A material's response to small but finite deformations can reveal the roots of its response to much larger deformations. Here, we identify commonalities in the responses of 2D soft jammed solids with different amounts of disorder. We cyclically shear the materials while tracking their constituent particles, in experiments that feature a stable population of repeated structural relaxations. Using bidisperse particle sizes creates a more amorphous material, while monodisperse sizes yield a more polycrystalline one. We find that the materials' responses are very similar, both at the macroscopic, mechanical level and in the microscopic motions of individual particles. However, both locally and in bulk, crystalline arrangements of particles are stiffer (greater elastic modulus) and less likely to rearrange. Our work supports the idea of a common description for the responses of a wide array of materials.

\end{abstract}

\maketitle

\section{Introduction}

Connecting a material's response under stress with its microscopic structure --- the arrangement of its constituent atoms or particles --- is a cardinal goal of materials science. In a crystalline solid this is done by accounting for the essential symmetries of the material, plus a relative handful of lattice defects. In amorphous or glassy materials, on the other hand, crystalline order may be nearly absent, and properties such as acoustic modes can be dramatically different~\cite{VanHecke:2010go,Anderson:1981fg}. Recently, Goodrich et al.~\cite{Goodrich:2014fl} proposed that these two cases delimit a continuum of disorder, but that over most of this continuum, a theory of amorphous solids is more useful to describe infinitesimal deformations and excitations. For instance, with just $\mathcal{O}(1\%)$ of particles deviating from crystalline order, the behaviour of the ratio of elastic to bulk modulus $G/B$ is much closer to that in a completely disordered system than to that in a perfect crystal.

Does this picture extend to finite deformations, ones large enough to rearrange particles? In amorphous solids, such rearrangements occur in localised groups of $\mathcal{O}(10)$ particles~\cite{Argon:1979iy,Schall:2007fd,Falk:2011co,Keim:2014hu}, often abstracted as shear transformation zones (STZs); these locations can be thought of as pre-existing packing defects, though not in any way so straightforward as in a crystal~\cite{Falk:2011co}. In recent experiments~\cite{Keim:2013je,Keim:2014hu}, we showed that under cyclic shear at finite strain amplitude $\gamma_0 \sim 3\%$, an amorphous solid reached a reversible plastic regime~\cite{Keim:2014hu} --- a steady state in which these regions would rearrange, and then reverse, on each cycle. Because it isolates a stable, limited population of structural relaxations, this kind of experiment suggests a way to compare the microscopic signatures of plasticity among different material structures. 

\begin{figure*}
\begin{center}
\includegraphics[width=4.5in]{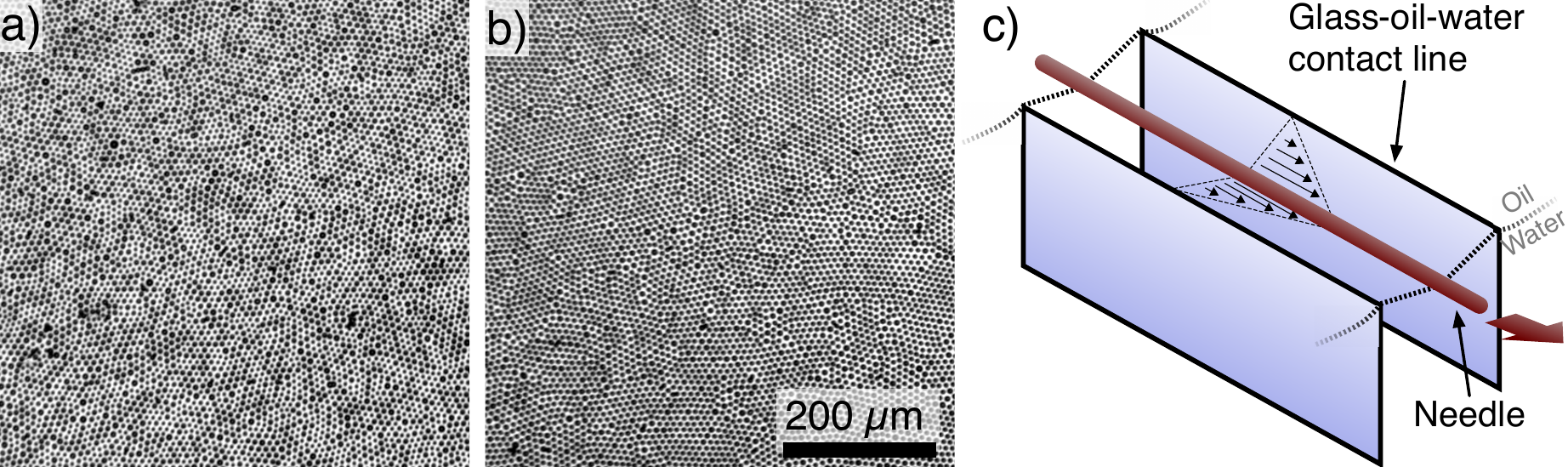}
\end{center}
\caption{Material and apparatus. Monolayers of \textbf{(a)}~bidisperse and \textbf{(b)}~monodisperse repulsive particles are adsorbed at a flat oil-water interface.
\textbf{(c)} Interfacial stress rheometer (ISR) apparatus. A magnetic force on the needle shears the monolayer uniformly (velocity profile sketched).
\label{fig:apparatus}}
\end{figure*}

Here, we observe the reversible plastic regime in two different packings that differ primarily in their amount of disorder, shown in Fig.~\ref{fig:apparatus}(a) and (b). The packing made with bidisperse particle sizes more closely resembles an amorphous material, while that made with monodisperse sizes more closely resembles a polycrystal. Our experiments combine tracking of many ($10^4$) individual particles with shear rheometry. We find that the monodisperse packing, with fewer disordered regions, also has fewer rearranging regions. However, the responses of each material are otherwise closely similar. Our results indicate that the conceptual picture proposed by Goodrich et al.\ may be extended to finite deformations, as also suggested by the recent simulations of Rottler et al.~\cite{Rottler:2014vj}. Our findings are also consistent in their broad outlines with prior studies involving steady shear~\cite{Shiba:2010uj, Biswas:2013jy, Katgert:2009eo}.

The model materials shown in Fig.~\ref{fig:apparatus}(a) and (b) are packings of polystyrene particles adsorbed at a water-decane interface~\cite{Keim:2013je,Keim:2014hu}. These particles have long-range electrostatic dipole repulsion~\cite{Masschaele:2010da}, so that without touching they form a stable jammed material. The bidisperse packing is a mixture of 4.1 and 5.6~$\mu$m-diameter particles, while the monodisperse packing includes the large species only. The materials are subjected to uniform shear deformations in a custom-made interfacial stress rheometer (ISR)~\cite{Brooks:1999ky,Reynaert:2008dm,Keim:2013je,Keim:2014hu}. As shown in Fig.~\ref{fig:apparatus}(c), a magnetised needle is embedded in the monolayer, parallel to the interface. The needle is centred between two vertical glass walls that form an open channel. Electromagnets centre the needle and drive it, applying a uniform shear stress $\sigma(t)$ on the material between the needle and the walls; thus, the rheometer is stress-controlled. Measuring the resulting strain $\gamma(t)$ allows us to compute material rheology. 
\section{Results}

In this section, we first provide global descriptions of the materials' particles and their rearrangements, followed by an analysis of how rearrangements are localised and organised.
\subsection{Material Composition and Confinement}

\begin{figure}
\begin{center}
\includegraphics[width=2.0in]{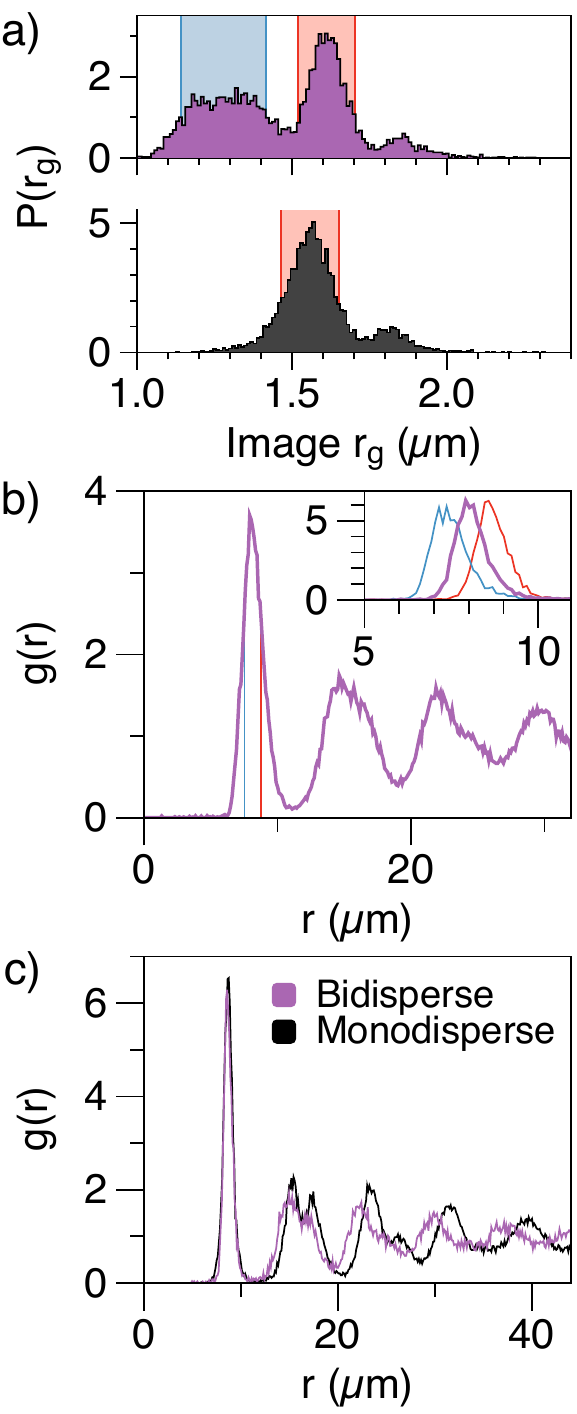}
\end{center}
\caption{Characterising particle packings. \textbf{(a)}~Distributions of apparent particle size (radius of gyration $r_g$) in images of bidisperse (top) and monodisperse (bottom) packings. Shaded rectangles above the curves define ``small'' particles (in bidisperse only) and ``large'' particles (in both packings).
\textbf{(b)}~Pair distribution function $g(r)$ of entire bidisperse packing, reflecting the separation of particles due to long-range repulsion. Inset: $g(r)$ first peaks of (left to right) small particles only, small-large pairs, large particles only. Small and large peak positions are indicated by vertical lines in main plot.
\textbf{(c)}~$g(r)$ among large particles within each packing. Packings were prepared to match first peak positions closely, indicating very similar confining pressure on these repulsive particles.
\label{fig:material}}
\end{figure}

Figure~\ref{fig:material}(a) shows the sizes of the solid particles in each packing, measured as the radius of gyration of the particle image (see Methods, below). While the large particles within each sample were drawn from the same stock, the details of image analysis were optimised separately for each packing, and so the $r_g$ of large particles differ slightly. Each packing also has a small population of abnormally large particles, which do not appear to play a disproportionate role in the dynamics.

Figures~\ref{fig:material}(b) and (c) use the radial pair correlation function $g(r)$ to describe the arrangement of particles due to their long-range repulsion. The first peak in $g(r)$ represents the typical spacing between pairs of neighbouring particles, which we denote as $a$. Figure~\ref{fig:material}(b) shows the overall $g(r)$ of the bidisperse packings; the inset shows the contribution of pairs of each species. The distribution of repulsion strengths within a species was previously studied by Park et al.~\cite{Park:2010fs}. Note that while the small and large species differ in solid size by $\sim 25\%$, their much greater \emph{effective} sizes differ by just $\sim 15\%$. 

We control number density, and hence repulsion strength and osmotic pressure, by changing the number of particles dispersed into the 6~cm-diameter experimental cell. Our materials may be considered athermal and jammed: we do not observe thermal motion, and strong long-range repulsions mean we are far beyond any jamming transition at which particles become under-constrained. Since the repulsive force between particles falls off monotonically with separation, interparticle spacings are a proxy for the pressure that confines the particles. Figure~\ref{fig:material}(c) shows that our samples are at very similar pressures: the separation of neighbouring large particles (almost all particles in the monodisperse packing) is the same in the monodisperse and bidisperse samples to within 1\%. Furthermore, the \emph{width} of the first peak is similar between packings, despite the greater heterogeneity in the environments of particles in the bidisperse packing. This suggests that the confining osmotic pressure is a good control parameter for interparticle interactions throughout the packing.
\subsection{Rheology and Rearrangements}

\begin{figure}
\begin{center}
\includegraphics[width=2.5in]{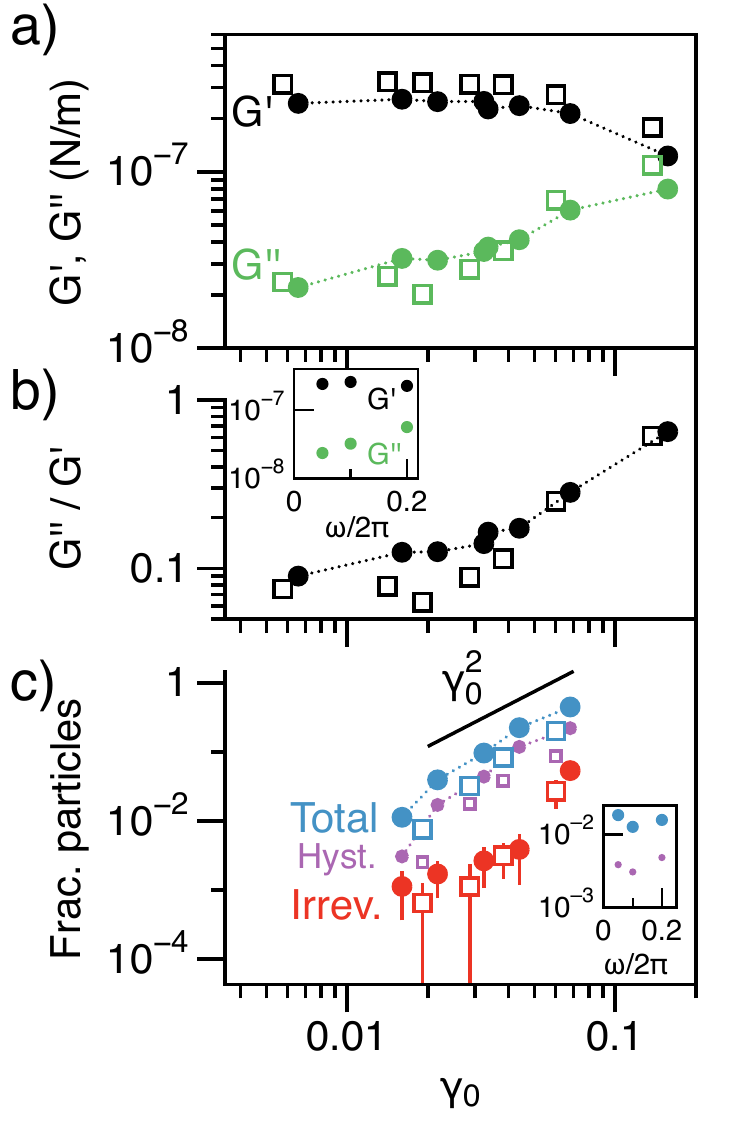}
\end{center}
\caption{Global measures of material response as a function of strain amplitude $\gamma_0$, for bidisperse (connected closed symbols) and monodisperse packings (open symbols). 
\textbf{(a)}~Oscillatory shear elastic ($G'$) and loss ($G''$) moduli are similar between the two packings.  
\textbf{(b)}~The ratio $G''/G'$ highlights the yielding transition. Inset: Loss modulus $G''$ depends weakly on driving frequency, as shown here for bidisperse packing at $\gamma_0 \approxeq 0.016$.
\textbf{(c)}~Rearranging fractions of particles in the steady state, measured as defined in text: total (peak-to-peak), hysteretic, and irreversible (stroboscopic). Error bars on the irreversible points represent standard deviation. $\gamma_0^2$ scaling is drawn for comparison. Total and hysteretic activity are much greater in the monodisperse packing. The onset of yielding corresponds to a sharp increase in irreversible rearrangement activity. Inset: total and hysteretic activity at $\gamma_0 \approxeq 0.016$ (here averaged over just 2--3 cycles) show weak dependence on frequency, with no clear trend.
\label{fig:microrheo}}
\end{figure}

Figure~\ref{fig:microrheo}(a) shows the oscillatory shear rheology of each material, as a function of strain amplitude $\gamma_0$ at 0.1~Hz. The materials have similar responses, with the monodisperse packing being stiffer (higher elastic modulus $G'$ at small $\gamma_0$). The inset of Fig.~\ref{fig:microrheo}(b) shows that dissipation at low $\gamma_0$ depends weakly on frequency~\cite{Keim:2013je}. As $\gamma_0$ is increased, each material goes through a rheological yielding transition, with $G'$ falling, and viscous dissipation rising (as measured by loss modulus $G''$). The yielding behaviour, summarised by the ratio $G'' / G'$ in Fig.~\ref{fig:microrheo}(b), is nearly identical for the two packings.

Simultaneously with shear rheometry, our experiments also track nearly all particles in a segment of the material. We use a long-distance microscope, a high-speed camera at 40 frames/s, and in-house freely-available particle-tracking~\cite{trackpyv02} and analysis~\cite{philatracks} software both to observe $\gamma(t)$ for rheometry, and to identify rearrangements among $\sim 4 \times 10^4$ particles~\cite{Keim:2013je,Keim:2014hu}. To identify rearrangements, we look for particle motions that are locally non-affine, as measured by the quantity $\dtwomin$~\cite{Falk:1998wm,Keim:2014hu}. We compute $\dtwomin$ between any times $t_1$ and $t_2$ by considering a particle and its two nearest ``shells'' of neighbours (within radius $\sim 2.5a$), and finding the best affine transformation that relates their positions at $t_1$ and $t_2$; $\dtwomin(t_1, t_2)$ is the mean squared residual displacement after subtracting this transformation, normalised by $a^2$. We define a rearranging particle as one with $\dtwomin(t_1, t_2) \ge 0.015$, a threshold comparable with one used for simulations of disordered solids~\cite{Falk:1998wm}. For an illustration of video microscopy, particle tracking, and $\dtwomin$, see Supplementary Movie 1.

Using these methods, we may measure the rate at which the material is altered by repeated cycles of driving. Before each experiment begins, we rejuvenate the material with 6 cycles of large-amplitude shearing ($\gamma_0 \sim 0.5$), then stop; once we resume shearing with a smaller amplitude, we observe that the rate of change decays during a transient and reaches a relatively steady value. The steady state appears to begin at roughly cycle 15 in each 30-cycle movie; this timescale varies little with $\gamma_0$, in contrast to the divergences reported in some studies of disordered solids~\cite{Regev:2013es,Fiocco:2013ds,Nagamanasa:2014jx} and sheared suspensions~\cite{Corte:2008tp}. The rate of irreversible rearrangements over time is plotted in Supplementary Fig.~1. That figure also shows the evolution of the monodisperse material at $\gamma_0 = 0.014$, which is not shown in the present results; that experiment showed significant irreversible change near the end of the recording, possibly due to an external disturbance.

Our analysis of rearrangements focuses on the steady state, and it attempts to capture all activity during a cycle: we detect both \emph{total} (peak-to-peak) rearrangements, due to the deformation between a minimum in strain $\gamma(t_\text{min})$ and the subsequent maximum $\gamma(t_\text{max})$; and \emph{irreversible} (stroboscopic) rearrangements that are the net result of the full cycle, as delimited by $(t_\text{min} + t_\text{max} \pm 2\pi \omega^{-1}) / 2$. In general, a particle in the total set of rearrangements, but not the irreversible set, is rearranging reversibly. Finally, we identify the sub-population of reversible rearrangements that are \emph{hysteretic}, requiring a buildup of stress to activate~\cite{Falk:1998wm,lundberg08,Falk:2011co,Keim:2014hu}. Within a cycle beginning at $t_\text{min}$, we obtain $t_\text{on}$ at the last video frame for which $\dtwomin(t_\text{min}, t) < 0.015$, and $t_\text{off}$ at the last frame with $\dtwomin(t_\text{min}, t) \ge 0.015$. These correspond to global strains $\gamma_\text{on}$ at which the particle rearranges during forward shear, and $\gamma_\text{off}$ at which it is reverted during reverse shear. We consider a rearrangement hysteretic if $\gamma_\text{on} - \gamma_\text{off}$ exceeds the largest strain increment $\Delta \gamma$ between video frames in the movie, ``on'' and ``off'' are in the first and second halves of the cycle respectively, and $\dtwomin \ge 0.015$ for at least 50\% of the intervening frames~\cite{Keim:2014hu}. The fraction of hysteretic particles is robust to changes in the $\Delta \gamma$ threshold: increasing the threshold by a factor of 10, or omitting it entirely, respectively decreases or increases the fraction by only $\sim 10\%$.  Supplementary Movie 1 features an example of a hysteretic rearrangement.

Figure~\ref{fig:microrheo}(c) shows the fraction of particles in these three rearranging populations, as a function of $\gamma_0$, averaged over the steady state. The measurements for the bidisperse packing are close to previously-published results for bidisperse packings at higher osmotic pressure~\cite{Keim:2014hu} (not shown). The present data show more rearrangements in the bidisperse packing than in the monodisperse packing, at all $\gamma_0$. However, the two packings' microscopic behaviours are otherwise strikingly similar. Reversible and hysteretic rearrangements occur in roughly the same proportion to each other, with a steep onset around $\gamma_0 = 0.02$, and thereafter scaling roughly as $\gamma_0^2$ even as the system yields. Irreversible activity, on the other hand, is minimal [$\mathcal{O}(10)$ particles in the field of view] and increases slowly below $\gamma_0 \sim 0.05$, at which point the system appears to undergo a microscopic yielding transition to a strongly irreversible steady state, as has been observed for other soft solids~\cite{Hebraud:1997ef,Petekidis:2002by,Hebraud:1997ef,Priezjev:2013hp,Keim:2013je,Regev:2013es,Schreck:2013hp,Fiocco:2013ds,Keim:2014hu,Nagamanasa:2014jx,Knowlton:2014ki}. We note that levels for the monodisperse packing at $\gamma_0 = 0.06$ are likely reduced because of a nascent slip layer near the wall; however, measurements of that experiment are consistent with trends at lower $\gamma_0$, and so we have included them here.
\subsection{Role of Disorder}

\begin{figure*}
\begin{center}
\includegraphics[width=6.5in]{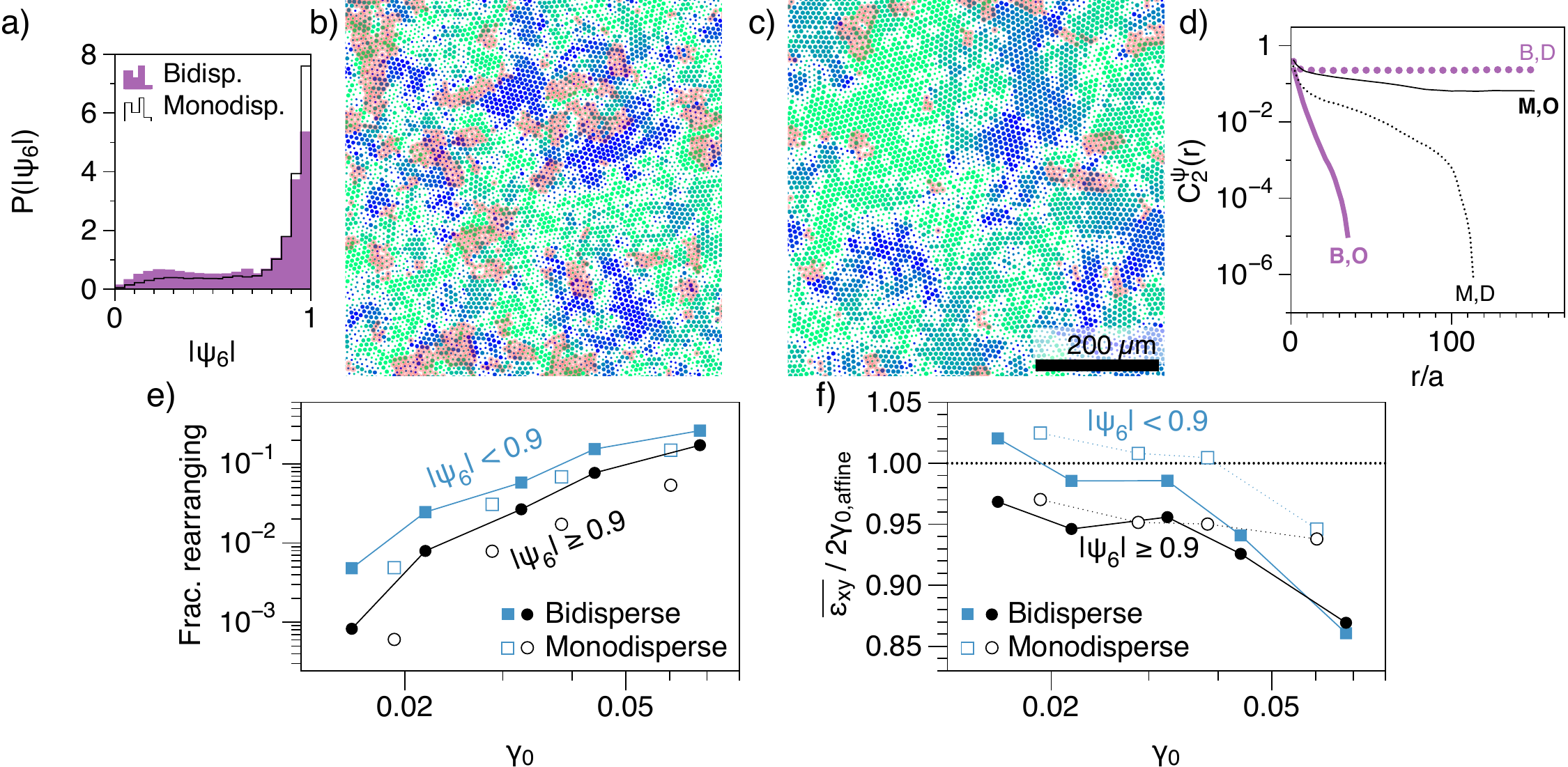}
\end{center}
\caption{Static crystalline structure and local deformation.
\textbf{(a)}~Magnitude distributions of bond order parameter $\psi_6$ in bidisperse (black line) and monodisperse (shaded region) packings.
\textbf{(b)}~$\psi_6$ of bidisperse packing region in Fig.~\ref{fig:apparatus}(a), in the steady state at $\gamma_0 = 0.044$. Dot size shows $|\psi_6| \ge 0.9$ (large) or $< 0.9$ (small); colour shows lattice director and is solely to guide the eye. Particles that rearrange are highlighted in red (see text).
\textbf{(c)} $\psi_6$ in monodisperse packing in Fig.~\ref{fig:apparatus}(b), at $\gamma_0 = 0.038$. There is a large difference in size and quality of crystalline grains.
\textbf{(d)} Two-point cluster function $C_2^\psi(r)$, the probability that two particles $r$ apart are in the same contiguous region, plotted for ordered ($|\psi_6| \ge 0.9$) and disordered ``phases'' within the bidisperse (labeled ``B,O''; ``B,D'') and monodisperse (``M,O''; ``M,D'') packings. The ordered phase of the monodisperse packing (``M,O'') and the disordered phase of the bidisperse packing (``B,D'') percolate their respective materials; their counterparts have limited range.
\textbf{(e)}~Fraction of particles with low ($\blacksquare$) or high ($\bullet$) local crystalline order $|\psi_6|$ that rearrange (total $\dtwomin \ge 0.015$), as a function of $\gamma_0$. Closed symbols: bidisperse; open symbols: monodisperse. Particles in disordered environments are more likely to participate in rearrangements, especially in monodisperse packings.
\textbf{(f)}~Local elastic deformation: elastic strain $\epsilon_{xy}$ near particles that do \emph{not} rearrange (total $\dtwomin < 0.015$) is normalised by global affine strain $2\gamma_{0, \text{affine}}$. Averages are plotted for particles with low and high $|\psi_6|$ as a function of $\gamma_0$. At low $\gamma_0$, low-$|\psi_6|$ particles appear ``softer'' (deformed by more than the global strain), and high-$|\psi_6|$ particles appear ``stiffer.'' At high $\gamma_0$, much of the global shear is accomplished by non-affine deformation (not plotted), and local structure becomes less relevant. 
\label{fig:stats}}
\end{figure*}

As noted above, the overall similarities in response between the two materials are despite their different distributions of effective particle sizes, which result in different propensities toward local crystalline ordering. To examine the packings' structures more closely, we compute the bond-order parameter $\psi_6$, identifying sixfold symmetry in the placement of a particle's neighbours:
\begin{equation}
\psi_6 = \frac{1}{N_r} \sum_{n=1}^{N_r} e^{i 6 \theta(\vec r_n - \vec r_0)}
\end{equation}
where $n$ runs over the $N_r$ neighbours that fall within $1.5a$ of the particle, and $\theta(\vec r_n - \vec r_0)$ denotes the angle that the vector from the particle to its neighbour makes with a fixed reference axis $\hat i$. The magnitude $|\psi_6|$ ranges from 0 to 1 and measures the degree to which the particle's neighbourhood resembles a hexagonal crystal, while the complex phase corresponds to the local lattice director. Figure~\ref{fig:stats}(a) shows the distribution of $|\psi_6|$ within each packing, showing that the bidisperse packing has significantly more particles with local disorder. 

The different prevalences of crystalline order are also clear within the portions of each packing shown in Figs.~\ref{fig:stats}(b) and (c), drawn to represent the $\psi_6$ of each particle. The monodisperse packing resembles a polycrystalline agglomeration of grains, each containing $\lesssim 100$ particles, whereas the analogous regions in the bidisperse packing are much smaller and contain numerous defects. We can quantify this structure, and any associated length scales, by computing a 2-point cluster function~\cite{Torquato:1988ek,Jiao:2009ft} for local order, $C_2^\psi(r)$. We define ordered and disordered phases using the threshold $|\psi_6| = 0.9$. Among particles of each phase we then identify contiguous clusters of particles. Defined generally, $C_2(r)$ is the probability that any two particles, separated by distance $r$, lie in the same contiguous cluster (see Methods section for details). It thus probes not only length scales, but also connectedness and percolation~\cite{Torquato:1988ek,Jiao:2009ft}. In Fig.~\ref{fig:stats}(d) we show $C_2^\psi(r)$ of the ordered and disordered phases within each packing. There are dramatic differences: in the bidisperse packing, the disordered phase effectively percolates the entire observed region (size $\sim 150a$), while the ordered phase has a characteristic length scale of $\sim 3$ particles. In the monodisperse packing, it is the ordered phase that percolates.  Here, extended networks of grain boundaries are suggested by the steep initial drop-off of the disordered-phase $C_2^\psi(r)$ (the grain boundary width) and shallower secondary decay.

The red shaded markers in Figs.~\ref{fig:stats}(b) and (c) indicate rearranging particles, ones with total $\dtwomin \ge 0.015$. Rearrangements tend to be localised to particles with low $|\psi_6|$~\cite{Keim:2014hu}. Similarly, Fig.~\ref{fig:stats}(e) shows that particles with $|\psi_6| < 0.9$ are more likely to rearrange than those with $|\psi_6| \ge 0.9$. This trend is especially pronounced in monodisperse packings.

Even when particles are \emph{not} rearranging, we find that their motion is correlated with $|\psi_6|$. We compute the local horizontal shear strain $\epsilon_{xy}$ by the same least-squares method as for $\dtwomin$, and normalise it by the least-squares shear strain of the entire packing, $\gamma_{0,\text{affine}} \simeq \gamma_0$. Figure~\ref{fig:stats}(f) shows that at low $\gamma_0$ (below yielding), $\epsilon_{xy}$ tends to be $\sim 10\%$ greater among particles with low $|\psi_6|$. The local $\epsilon_{xy}$ of a specific region is due both to its local elastic moduli and to the local stress on that region, and it is difficult to separate their respective contributions. However, we may expect the local elastic moduli to be most influenced by local structure, while stress would be heavily influenced by the properties and deformations \emph{elsewhere} in the material, to preserve force balance.  The correlation of local $\epsilon_{xy}$ with local $|\psi_6|$ therefore suggests that disordered regions themselves are at least slightly ``softer,'' even when they do not rearrange.

As $\gamma_0$ is increased in Fig.~\ref{fig:stats}(f), we see the effects of \emph{non}-affine deformation. Rearrangements locally accomplish shear deformation by a dissipative plastic process, rather than by accumulating elastic stress; they reduce bulk stress at a given strain. The result is that the non-rearranging portions of the system are under less stress, and deform less, than they would if no rearrangements happened. This is evidenced by the downward trend of $\epsilon_{xy}/2\gamma_{0,\text{affine}}$ in Fig.~\ref{fig:stats}(f). Additionally, the distinction between low- and high-$|\psi_6|$ particles appears to vanish at large $\gamma_0$, in contrast to the clear difference at small $\gamma_0$. The meaning of this change is unclear.
\subsection{Spatial Organisation}

\begin{figure}
\begin{center}
\includegraphics[width=2.5in]{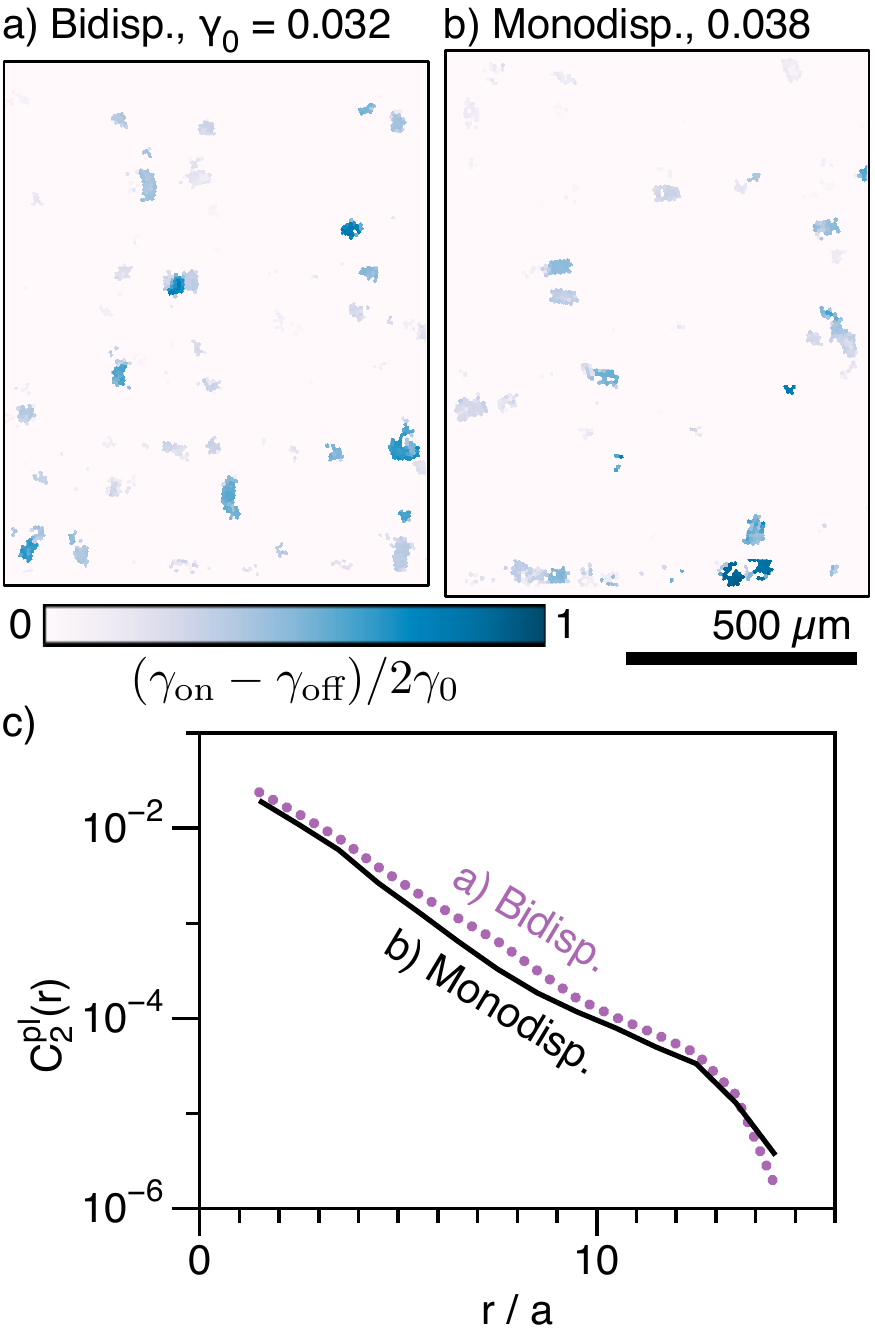}
\end{center}
\caption{Material structure has little effect on spatial organisation of rearrangements.
\textbf{(a)}~Map of normalised particle hysteresis, $(\gamma_\text{on} - \gamma_\text{off}) / 2\gamma_0$, for one steady-state shear cycle of the bidisperse packing at $\gamma_0 = 0.032$. A segment between the wall (bottom edge) and the needle (top edge) is shown. 
\textbf{(b)}~Similar map for monodisperse packing, $\gamma_0 = 0.038$. Although $\gamma_0$ is slightly \emph{higher} than in (a), there are \emph{fewer} rearranging regions, consistent with Fig.~\ref{fig:microrheo}(c). Note that $\gamma_\text{on} - \gamma_\text{off}$ is nearly uniform within each cluster, suggesting that particles rearrange and reverse cooperatively. 
\textbf{(c)}~Two-point cluster function $C_2^\text{pl}(r)$ among clusters of hysteretic particles (see text), for the data represented in (a) and (b). Radius $r$ is normalised by typical particle spacing $a$. The last point of each curve indicates the maximal diameter of the largest cluster(s). The shape and size of clusters are similar between the two packings.
\label{fig:spatial}}
\end{figure}

As discussed above, plastic (i.e.\ hysteretic) rearrangements tend to involve discrete clusters of particles, which may be described as shear transformation zones (STZs)~\cite{Falk:2011co,Keim:2014hu}.  In our discussion of Fig.~\ref{fig:microrheo}(c), we noted that the number of hysteretically-rearranging particles is 2--3 times greater in the bidisperse packing than in the monodisperse packing, at similar $\gamma_0$. The results in Fig.~\ref{fig:spatial} suggest that we should interpret this difference in terms of the \emph{number} of rearranging clusters, not their shape or size. We colour each particle according to the extent to which its rearrangement is hysteretic, $\gamma_\text{on} - \gamma_\text{off}$. Visual examination reveals that clusters have comparable size in the two packings. Figure~\ref{fig:spatial}(c) shows this more quantitatively using a 2-point cluster function~\cite{Torquato:1988ek,Jiao:2009ft} for these reversible plastic regions, $C_2^\text{pl}(r)$, which also reveals the maximum cluster diameter, given by the largest $r$ for which $C_2^\text{pl}(r) > 0$. We find this broad similarity throughout the reversible plastic regime of $0.02 \lesssim \gamma_0 \lesssim 0.04$, suggesting that the primary result of changing $\gamma_0$ or microscopic structure is to change the number and placement of these regions.
\section{Discussion}

The comparisons we have presented are significant in part because our experiments allow us to prepare differently-structured packings confined at similar osmotic pressure. We argue that pressure is likely the best control parameter for this system: other control parameters such as area fraction, particle overlap, and coordination number are difficult to define usefully in the presence of long-ranged repulsions, and area fraction has a different mechanical interpretation in ordered and disordered packings~\cite{VanHecke:2010go}.

At the smallest $\gamma_0$ we measure a $G'$ for the monodisperse packing that is $\sim 50\%$ greater [Fig.~\ref{fig:microrheo}(b)]. Perhaps more importantly, this discrepancy appears \emph{within} the packings: even as the entire packing deforms nearly elastically, regions with local crystalline order are deformed less than disordered regions [Fig.~\ref{fig:stats}(f)].

At larger $\gamma_0$, our experiments allow a more detailed comparison of the mechanical properties of each packing, because of the observations afforded by the reversible plastic regime. The observed regions of hysteretic rearrangements lend themselves to a theoretical description of material deformation, for the same reasons that favour shear transformation zones (STZs)~\cite{Falk:2011co}: they dissipate energy and change bulk rheology by their hysteresis; they may be considered as two-state subsystems with long-range interactions; and they arise out of the static microstructure of the material. In such a model, the material's response to large-amplitude or steady shear is built up out of many consecutive rearrangements, drawn from a continually-replenished population of STZs~\cite{Falk:2011co}. Observations of the reversible plastic regime thus connect static structure with yielding behaviour and steady shear, both in amorphous solids, and perhaps in polycrystals~\cite{Shiba:2010uj,Biswas:2013jy,Tamborini:2014kk}.

Our results indicate that the reversible plastic regime, and the type of rearranging regions it reveals, are shared by more- and less-disordered packings. This in turn hints at a common understanding of the localisation of rearrangements in a range of materials. The idea of a common understanding is also supported by recent simulations by Rottler et al.~\cite{Rottler:2014vj}, that suggest that a method to predict localisation in bulk disordered packings also works at lattice defects and grain boundaries. Notably, our observations show anecdotally that rearrangements tend \emph{not} to occur at lattice defects within crystallites, and are instead favoured at grain boundaries, as illustrated by Fig.~\ref{fig:stats}(c).

Our work also complements comparative studies of solids under \emph{steady} shear. Experiments with steadily-sheared foams have shown that a small amount of disorder makes the flowing material's rheology qualitatively like that of an amorphous solid~\cite{Katgert:2009eo}. Furthermore, simulations of steady shear suggest that as in amorphous solids, particles in polycrystals rearrange with spatiotemporal heterogeneity~\cite{Biswas:2013jy} and with similar sliding of particles past one another~\cite{Shiba:2010uj}. In a notable difference from our results, however, in simulations by Shiba et al.~\cite{Shiba:2010uj} the scale of localisation in polycrystals may be much larger because of extended grain boundaries; for oscillatory shear we see no such trend (Fig.~\ref{fig:spatial}[c]).

\section{Conclusions}

In this paper, we have investigated the response of materials to small but finite-amplitude deformations, considering how this response depends on the degree of disorder in the materials' microscopic structure. We characterised the materials' steady-state response to cyclic shear, which can be controlled by a stable population of rearrangements that occur and reverse on each cycle. Through experiments on packings of particles with monodisperse and bidisperse sizes at the same confining pressure, we have shown that despite clear differences in material structure, the response is very similar, in both macroscopic rheology and microscopic particle motions. A reversible plastic steady state arises in each material for a range of strain amplitudes $0.02 \lesssim \gamma_0 \lesssim 0.05$, rearrangements occur in clusters of $\mathcal{O}(10)$ particles, and these clusters are correlated with more-disordered regions of the material. We find just two major differences: at small amplitudes, a higher apparent elastic modulus for particles with crystalline order, both locally (Fig.~\ref{fig:stats}[f]) and at the bulk scale (Fig.~\ref{fig:microrheo}[a]); and at larger amplitudes, a 2--3-fold greater population of rearrangements in the more-disordered material (Fig.~\ref{fig:microrheo}[c]), occurring in clusters of similar length-scale (Fig.~\ref{fig:spatial}).  The consequences of this population difference remain an open question, and could include the details of rheology, the precise onset of irreversibility in the steady state, and the materials' memory capacity~\cite{Keim:2011dv,Fiocco:2014bz}.

This work contributes to an understanding of material plasticity and mechanical response as situated along a continuum of disorder, from perfect crystals to maximally random packings~\cite{OHern:2003gu}. Our experiments add to the evidence that a common set of tools might be used to describe, and someday predict, the behaviour of a vast array of particulate solids with nearly any amount of disorder.
\section{Experimental Methods}

\subsection{Materials and Sample Preparation}

Particles are sulphate latex (Invitrogen), with nominal diameters 4 and 6~$\mu$m. To prepare samples, we rinse and sonicate particles 4 times in deionised water, which yields more uniform interparticle forces~\cite{Park:2010fs}. They are then resuspended in a water-ethanol mixture (50\% by volume), so that the suspension is violently dispersed when added to the interface. To reduce polar contamination that could screen repulsion, the decane superphase (``99+\%,'' Acros Organics) is treated with activated aluminium oxide powder (Alfa Aesar), which is then removed by filtration (Qualitative No.~1, Whatman). The apparatus is cleaned before each experiment by repeated sonication and rinsing in deionised water and ethanol.

\subsection{Particle sizes}

The image radius of gyration $r_g$ of a particle is computed as
\begin{equation}
r_g \simeq \Bigg( \frac{ \int_0^R r^2 I(r, \theta)\ r\ dr\ d\theta }
{ \int_0^R I(r, \theta)\ r\ dr\ d\theta} \Bigg)^{1/2}
\end{equation}
where ``$\simeq$'' represents a discrete (pixel-wise) approximation, and $I(r, \theta)$ is the image intensit, at distance $r$ and polar angle $\theta$ relative to the particle centroid. The integral is performed over the portion of the image, with radius $R$, that encompasses the particle~\cite{Crocker:1996wp,trackpyv02}.

\subsection{Quasistatic and 2D Assumptions}

For the material and our rheometry to be effectively 2-dimensional, the boundary conditions in the third dimension must be approximately stress-free. This is quantified by the Boussinesq number $Bq = |\eta^*| a / \eta_l$, where $\eta^*$ is the material's observed complex viscosity, $a = 230$~$\mu$m is the needle diameter, and $\eta_l \simeq 10^{-3}$~Pa~s is the oil and water viscosity~\cite{Reynaert:2008dm}.  Here, $Bq \sim 10^2$, so that typical stresses within the plane are much stronger than viscous drag from the liquid bath. Our experiments may also be considered quasistatic, insofar as individual rearrangements occur on a timescale ($\lesssim 1$~s) much shorter than a period of driving or the largest inverse strain rate (both $\sim 10$~s). Consistent with this assumption, we observe a weak dependence of rearrangement activity on frequency (Inset of Fig.~\ref{fig:microrheo}[c]); other, longer relaxation timescales may exist in the system, but we do not observe their influence.

\subsection{Removal of Spurious Rearrangements}

A temporary mis-identification of the particles can create the appearance of a rearrangement. Most such errors involve the abrupt, momentary displacement of a single, isolated particle by a large fraction of the interparticle spacing $a$. Genuine rearrangements, by contrast, involve the motion of many nearby particles, especially when they generate $\dtwomin$ far above the threshold of 0.015. When counting rearranging particles for Fig.~\ref{fig:microrheo}(c), we therefore discard a particle if its $\dtwomin$ exceeds the median of its neighbours' by more than 0.2. This step nearly eliminates spurious rearrangements without altering genuine ones.

Because our apparatus features a long-distance microscope (Infinity K2/SC) suspended over the sample and magnetic coils, the microscope is more sensitive to high-frequency transient vibrations than the sample itself. During the experiments with the monodisperse packings, such vibrations intermittently made many particles blurred and indistinct. When the transient lasted for just a few video frames, the particle-tracking algorithm~\cite{trackpyv02} allowed us to safely discard those frames; when the transient was more severe, we discarded the entire cycle, as evidenced in Supplementary Fig.~1.

\subsection{Computing $C_2^\psi(r)$ and $C_2^\text{pl}(r)$}

Particles of interest are selected by their hysteretic rearrangements [for $C_2^\text{pl}(r)$] or local sixfold symmetry $|\psi_6|$ [for $C_2^\psi(r)$]. Among these particles we find clusters (in graph theory, connected components) that are connected topologically via nearest-neighbour relationships (i.e.\ separated by $\le 1.5a$). To avoid the most tenuous clusters, we require that every cluster can survive the removal of any one particle~\cite{Hagberg:2008tk}. $C_2(r)$ is then defined as $\sum_i N_P^i(r) / N_S(r)$, where $N_S(r)$ is the number of interparticle pair distances of length $r$ in the entire packing, and $N_P^i(r)$ is the number of pair distances of length $r$ within the cluster $i$~\cite{Torquato:1988ek,Jiao:2009ft}. 

\vskip 3em

\section{Acknowledgements}

We thank Denis Bartolo, Ludovic Berthier, Alison Koser, Carl Goodrich, Andrea Liu, Sal Torquato, and Martin van Hecke for helpful discussions. This work was supported by the Penn NSF MRSEC (DMR-1120901).

\bibliographystyle{naturemag}
\bibliography{references,references-suppmat}

\end{document}